\documentclass[fleqn,twoside]{article}
\usepackage{espcrc2}
\usepackage{graphicx}
\title{Dynamical suppression of large instantons%
\thanks{Talk presented by G. M\"unster}}

\author{Gernot M\"unster\address{Institut f\"ur Theoretische Physik,
             Universit\"at M\"unster,
             Wilhelm-Klemm-Str.~9, D-48149 M\"unster, Germany,
             e-mail: munsteg@uni-muenster.de}
        and
        Christel Kamp\address{Institut f\"ur Theoretische Physik und
             Astrophysik, Universit\"at Kiel,
             Leibnizstr.~15, D-24098 Kiel, Germany,
             e-mail: kamp@theo-physik.uni-kiel.de}}

\begin{document}

\begin{abstract}
We investigate the distribution of instanton sizes in the framework of a
simplified model for ensembles of instantons.  This model takes into
account the non-diluteness of instantons.  The infrared problem for the
integration over instanton sizes is dealt with in a self-consistent
manner by approximating instanton interactions by a repulsive hard core
potential.  This leads to a dynamical suppression of large instantons.
The characteristic features of the instanton size distribution are
studied by means of analytic and Monte Carlo methods.
We find a power law behaviour for small sizes, consistent with the
semi-classical results.
At large instanton sizes the distribution decays exponentially.  The
results are compared with those from lattice simulations.
\vspace{1pc}
\end{abstract}

\maketitle
%
%
\section{INTRODUCTION}
Instantons are field configurations of non-abelian SU($N$) gauge
theories, which lead to non-perturbative effects. In recent years they have
been studied in lattice gauge theories by means of Monte Carlo calculations
by different groups \cite{FGPS96,FGPS97,Smith,Hasenfratz}.

In the dilute gas approximation the logarithm of the partition function
contains an integral over instanton sizes,
$\int\!\!d\rho \, \rho^{-5} (\rho \Lambda)^{b}$,
where $b = 11 N / 3$
and $\Lambda$ is the scale parameter. The
integrand increases with $\rho$, leading to an infrared divergence.
This is an artifact of
using the semiclassical approximation after it has become invalid, i.e.\ for
$\rho \geq 1/\Lambda$. If the semiclassical approximation is meaningful at
all, a solution of this problem in the context of the full instanton
ensemble is required.

The simplest way is to cut the integrations off at some ad-hoc value
$\rho_c$.  But the dominant contribution comes from large $\rho_j$ near the
cut-off where the assumption of diluteness fails.  Moreover, the introduction
of an ad-hoc cut-off leads to inconsistencies with the renormalization group
\cite{IMP}.

In order to solve the problem it has
been proposed that instanton sizes are cut off in a dynamical way
\cite{IMP,Mue82}.  The cut-off should originate from
configurations where instantons start to overlap.
The interaction is expected to suppress overlapping
instantons and to result in a self-consistent cut-off.

In connection with the dynamical cut-off the distribution of instanton sizes
is of central importance.
For small sizes the distribution is predicted to be
\begin{equation}
\label{smallrho}
n(\rho) \sim \rho^{b-5}
\end{equation}
by the dilute gas approximation. For large sizes $\rho$, where the dynamical
cut-off is in effect, not much is known about the distribution.  There are
arguments \cite{IMP,Dyakonov,Shuryak99} in favour of a suppression like
\begin{equation}
\label{largerho}
n(\rho) \sim \exp (-c \rho^p) \qquad \mbox{with} \quad p = 2 \,.
\end{equation}

We have investigated \cite{Kamp} the distribution of instanton sizes in
a model \cite{Mue82} where the instanton interactions are approximated by a
repulsive hard core potential. The radius of an instanton core
varies proportional to the size $\rho_j$ of the instanton.
Although this approximation appears
to be crude, the general features of the instanton ensemble with a dynamical
cut-off are present.
%
%
\section{SIZE DISTRIBUTION}

In $d=1$ dimensions the distribution of instanton sizes is exactly given by
\begin{equation}
\label{nrho1}
n(\rho) = \frac{C}{b} \, \rho^{b - 2} \, \mathrm{e}^{- c \rho} \,.
\end{equation}
where we recognize an exponential suppression of large instanton sizes.

In higher dimensions, $d>1$, we obtained approximate expressions
by means of a van der Waals type approximation \cite{Mue82},
\begin{equation}\label{nrho}
n(\rho) = \frac{C d}{b} \, \rho^{b-d-1} \, \exp (- c \rho^{d}) \,.
\end{equation}
For small $\rho$ it grows powerlike with the semiclassical exponent
$\alpha = b - d - 1$.
The value of the exponent $p = d$ in the exponential
decay at large $\rho$ should be considered with reservations, because
the saddle point approximations are of uncertain quality there.

We have studied the size distribution also by means of
grand canonical Monte Carlo
simulations of the simplified instanton gas model.
In the case of $d=1$ dimensions the Monte Carlo data agree very well with
the available exact result (\ref{nrho1}).

\begin{figure}[ht!]
\includegraphics[width=5.0cm,angle=270]{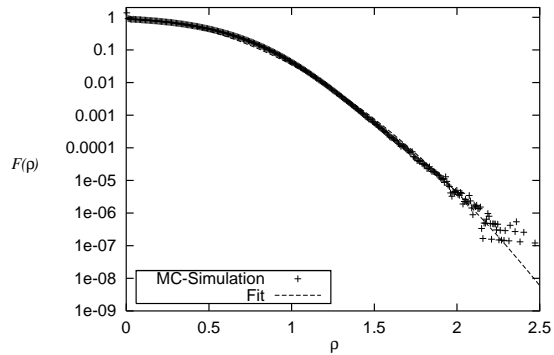}
\vspace*{-10mm}
\caption{\rm $F(\rho)$ in $d=4$ dimensions from a Monte Carlo
simulation in comparison with the fit.}
\label{fit}
\vspace*{-8mm}
\end{figure}

In the more interesting case of four space-time
dimensions ($d=4$) we consider
$\alpha = 7/3$, which is the value for SU(2) gauge theory.
The resulting size distribution shows the expected behaviour.
For small instanton radii a power law with exponent $\alpha$ can be confirmed.
In order to study the behaviour of $n(\rho)$ for large $\rho$ we
considered the ratio
$F(\rho) = n(\rho) / \rho^{\alpha}$
and tried fits of the form
$F_{fit}(\rho) = a \exp (-c \rho^p)$.
In agreement with the theoretical results they showed that $c$ depends on
$\alpha$, while $p$ is nearly independent of it. In Fig.~\ref{fit} the
result of a fit in the interval $[0,2.25]$ is shown.

The main interest is in the exponent $p$. We find $a\approx 0.89$, and the
fit leads to $c=3.3 \pm 0.2$ and $p=1.9 \pm 0.2$.  For $\alpha=6$, the SU(3)
case, the results for $p$ are the same within the present errors.

In recent years much effort has been devoted to lattice Monte Carlo
calculations of properties of the instanton ensemble, and some quantitative
statements have been given.
For small $\rho$, lattice
calculations appear to support the power law (\ref{smallrho}).
For the
large-$\rho$ distribution, de Forcrand et al.\ predict an exponential
decrease with $p = 3 \pm 1$ from their $SU(2)$ lattice data \cite{FGPS97}. 
In contrast to this, Smith and Teper conclude form their $SU(3)$ simulations
a decay according to
$\rho^{-\xi}$ with $\xi \approx 10 \ldots 12$ \cite{Smith}.

To conclude,
fits to our numerical Monte Carlo results suggest a behaviour like
\begin{equation}
n(\rho) \stackrel{\rho\to\infty}{\sim}
\rho^{\alpha} \, \exp (-c \rho^2) \,.
\end{equation}

The results indicate that our simplified model reproduces the main
features of instanton ensembles with a dynamical infrared cut-off.
%
%

%

\begin{thebibliography}{99}
\bibitem{FGPS96}
Ph.~de Forcrand, M.~Garc{\'\i}a P\'erez and I.-O.~Stamatescu,
Nucl.\ Phys.\ B (Proc.\ Suppl.) 47 (1996) 777.
%
\bibitem{FGPS97}
Ph.~de Forcrand, M.~Garc{\'\i}a P\'erez and I.-O.~Stamatescu,
Nucl.\ Phys.\ B\,499 (1997) 409.
%
\bibitem{Smith}
D.A.~Smith and M.J.~Teper,
Phys.\ Rev.\ D\,58 (1998) 014505.
%
\bibitem{Hasenfratz}
A.~Hasenfratz and C.~Nieter,
Phys.\ Letters B\,439 (1998) 366.
%
\bibitem{IMP}
E.-M.~Ilgenfritz and M.~M\"uller-Preu{\ss}ker,
Nucl.\ Phys.\ B\,184 (1981) 443.
%
\bibitem{Mue82}
G.~M\"unster,
Z.\ Phys.\ C, Particles and Fields 12 (1982) 43.
%
\bibitem{Dyakonov}
D.I.~Dyakonov and V.Y.~Petrov,
Nucl.\ Phys.\ B\,245 (1984) 259.
%
\bibitem{Shuryak99}
E.V.~Shuryak,
\texttt{hep-ph/9909458}.
%
\bibitem{Kamp}
C.~Kamp, G.~M\"unster,
Eur.\ Phys.\ J.\ C\,17 (2000) 447, 
\texttt{hep-th/0005084}.
%
\end{thebibliography}
\end{document}